\documentclass[iop]{emulateapj}
\usepackage{amsmath}
\usepackage{natbib}
\usepackage{xspace}
\usepackage{subfigure}

\renewcommand{\deg}{^{\circ}}
\newcommand{\fov}{\textsc{fov}\xspace}
\newcommand{\lsa}{\textsc{lsa}\xspace}
\newcommand{\msa}{\textsc{msa}\xspace}
\newcommand{\tam}{\textsc{tam}\xspace}
\newcommand{\tams}{\textsc{tam}s\xspace}
\newcommand{\eas}{\textsc{eas}\xspace}
\newcommand{\CR}{\textsc{cr}\xspace}
\newcommand{\CRs}{\textsc{cr}s\xspace}
\newcommand{\de}{\textrm{d}}
\newcommand{\dec}{\textsc{dec.}\xspace}
\newcommand{\argo}{\textsc{argo-ybj}\xspace}
\newcommand{\ra}{\textsc{r.a.}\xspace}

\newcommand{\etal}{{et al.}}

\begin{document}
\title{Time-average based methods for multi-angular scale analysis of cosmic-ray data}

\shorttitle{Time-average methods}
\shortauthors{Iuppa \etal}

\author{R.~Iuppa\altaffilmark{1,2} \& G.~Di Sciascio\altaffilmark{2}}
\email{roberto.iuppa@roma2.infn.it}
\email{giuseppe.disciascio@roma2.infn.it}
\altaffiltext{1}{Dipartimento di Fisica dell'Universit\`a ``Tor Vergata''
                  di Roma, via della
                  Ricerca Scientifica 1, 00133 Roma, Italy.}
\altaffiltext{2}{Istituto Nazionale di Fisica Nucleare, Sezione di
                  Roma Tor Vergata, via della Ricerca Scientifica 1, 00133 Roma, Italy.}


\begin{abstract}
In the last decade, a number of experiments dealt with the problem of measuring the arrival direction distribution of cosmic rays, looking for information on the propagation mechanisms and the identification of their sources. Any deviation from the isotropy may be regarded to as a signature of unforeseen or unknown phenomena, mostly if well localized in the sky and occurring at low rigidity. It induced experimenters to search for excesses down to angular scale as narrow as $10\deg$, disclosing the issue of properly filtering contributions from wider structures.
A solution commonly envisaged in these years based on time-average methods to determine the reference value of cosmic ray flux. Such techniques are nearly insensitive to signals wider than the time-window in use, thus allowing to focus the analysis on medium- and small-scale signals. Nonetheless, often the signal cannot be excluded in the calculation of the reference value, what induce systematic errors.
The use of time-average methods recently brought to important discoveries about the medium-scale cosmic ray anisotropy, present both in the northern and southern hemisphere.
It is known that the excess (or the deficit) is observed as less intense than in reality and that fake deficit zones are rendered around true excesses, because of the absolute lack of knowledge \emph{a-priori} of which signal is true and which is not.
This work is an attempt to critically review the use of time average-based methods for observing extended features in the cosmic-ray arrival distribution pattern.
%
\end{abstract}
\section*{Introduction}
Large field of view (\fov) experiments operated for cosmic-ray (\CR) physics collect huge amount of high-quality data, making possible to study the \CR arrival distribution with remarkable detail. Either satellite-borne or ground-based detectors are considered, many collaborations coped with the measurement of the \CR intensity all over the portion of the sky they observed. They all looked for deviations from the isotropic distribution, as any signature of anisotropy provides essential information on \CRs and the medium that they propagate through.

Apart from the search for gamma-ray emission from point-like (or \emph{quasi} point-like) sources, either in the MeV energy range on-board satellites and in the TeV region with ground-based telescopes, directional data are analyzed to map the \CR gradient all over the sky at every angular scale.
\vspace{3mm}\\
\textbf{Signal as deviation from the isotropy}\\
Any motion of the laboratory system with respect to the \CR plasma turns to a dipolar signature with a maximum in the direction of the motion. This is true for any ``still system'' we want to consider: it might be well the Solar System (i.e. the motion of the Earth around the Sun is factorized) or the Galaxy itself (the motion of the Solar System around the Galaxy center is factorized). Such a process of dipole generation is commonly referred to as ``Compton-Getting'' effect \citep{ComptonGetting} and was observed by a number of experiments \citep{EasTop1996,Tibet2004,Tibet2007,Icecube2011}. In this last case, as the amplitude and the phase of the signal are analytically predictable, the observation is commonly considered as the starting point of any anisotropy analysis, as it demonstrates the reliability of the detector and the analysis methods to be fine-tuned. Concerning the ``galactic'' Compton-Getting effect, the importance of this measurement lays in determining whether the \CR plasma is co-moving or not with the ``still system'' under study \citep{Tibet2006}.

Moving to narrower scales, it is known that a \CR ``pure'' anisotropy, i.e. not due to expected Compton-Getting effects, exists down to angular scales as wide as $\sim60\deg$. It was observed by ground-based detectors ever since 1930s and most recent experiments represented it in 2D sky-maps (see \citealt{review2012} and references therein). Such a ``large-scale'' anisotropy (\lsa) is of fundamental importance, commonly interpreted as a signature of the propagation of \CRs in the local medium.

Being charged particles, \CRs have trajectories deflected by magnetic fields,
so that their rigidity sets up a ``magnetic horizon'', i.e. a distance below which the observed arrival distribution contains information about the interaction of \CRs with the medium that they propagate through. The diffusion approximation effectively explains the observations beyond this horizon.
Thus, GeV-TeV \CRs are an effective tool to probe magnetic fields within the Solar System (up to the Heliotail) \citep{LazarianDesiati}. The multi-TeV region is important to study the \CR propagation in the Local Inter-Stellar Medium (LISM), whereas higher energy \CRs may reveal important features of the galactic magnetic fields. If electrons ($e\pm$) are considered, synchrotron energy losses should be accounted for in defining the magnetic horizon (remarkably closer than for protons of the same rigidity). Apart from that, the line is the same and gives the importance of any attempt to measure the anisotropy even in the $e\pm$ channel. The \CR electron anisotropy was recently searched for by the Fermi experiment \citep{FermiElectrons}, though with null result.

Since 2009 ``medium-scale'' anisotropy (\msa) structures were observed in the \CR distribution down to angular scale as wide as $10\deg$, their origin still unexplained (\CR source region, magnetic structures focusing \CRs, etc\dots) \citep{Tibet2007, MilagroMSA, ARGOMSA}. They were observed both in the Northern and the Southern hemisphere at TeV energy and they do not appear correlated to each other. It is quite natural to expect that a well defined \CR pattern at any angular scale will be found as soon as statistics will be enough. The narrower the structures the closer their origin, what explains the growing interest towards this phenomenon \citep{LazarianDesiati, DruryAharonian, SalvatiSacco, GiacintiSigl}.

Besides these \CR signals, diffuse gamma-ray emission is often measured by satellite experiments in the GeV range \citep{FermiDiffuse} and extensive air-shower (EAS) arrays observed the diffuse emission from the Galactic Plane at TeV energy \citep{MilagroDiffuse}. Structures as wide as $10\deg-20\deg$ have to be properly extracted from the background (the overwhelming \CR contribution or the average photon content) and the analysis methods used to do that often rest on the same ideas exploited to measure the \CR anisotropy.
\vspace{3mm}\\
\textbf{Detection techniques}\\
Either the \CR anisotropy or the diffuse emission from the Galactic plane is considered, the experimental issue of properly detecting and estimating the intensity of a signal as bright as $10^{-4}-10^{-3}$ with respect to the average isotropic flux of \CRs has to be dealt with.

It translates in estimating the exposure of the detector with accuracy well below that threshold, to avoid that detector effects mimic signals due to physics. To keep the exposure map under control down to $\sim10^{-4}$ is a challenge even for the most stable experiment, as unavoidable changes in the operating conditions occur and not always on line corrections can be readily applied. Both for satellite experiments and ground-based detectors the envisaged solution is to estimate off-line the exposure, relying on statistical methods applied to the large data-set available.

Good results can be achieved by combining Monte Carlo simulations and the record of the operating conditions (see for instance \citep{FermiDiffuse}). Otherwise, if simulations do not reach the needed accuracy, like for EAS experiments\footnote{The important effect of temperature and pressure variations on the atmospheric depth and, consequently, on the trigger efficiency of EAS arrays cannot be taken into account down to $10^{-4}$-$10^{-3}$}, or just to have a simulation-independent result, the exposure is estimated from data themselves, by exploiting some (assumed) symmetries in the data acquisition.

A number of data-driven methods to estimate the exposure exist, although all of them are based on geometrical properties of the detector acceptance (see for instance the ``equi-zenith'' methods \citep{Tibet2005}) and/or on the uniformity of the trigger rate within a certain period \citep{Alexandreas, Fleysher}. As well pointed out by \citep{Fleysher}, these symmetries are \emph{assumed} to be valid in some conditions and such an assumption is part of the null-hypothesis against which signals are tested.

Among all the techniques that experimenters developed to estimate the exposure, this paper is focused on those based on the time-average of collected data. Whatever its particular implementation is, any time-average method (\tam) relies on the assumption that the signal to be interpreted as background can be filtered out by averaging the event rate in a certain time window. The time-average can be performed directly, i.e. by integrating the event rate within the time-window and then dividing by the window width (\emph{direct integration} method), or by using Monte Carlo techniques. In the latest case, each event is associated with a number $N$ of ``fake'' events having all the same experimental characteristics but different arrival time, sampled according to the measured trigger rate. After ``time swapping'' (or ``shuffling'' or ``scrambling''), the over-sampling factor $N$ is accounted for and the final result makes this approach equivalent to the Direct Integration method. Actually, the only difference is that the integration is performed via Monte Carlo instead of directly.

The use of \tams is favored by the property well known in signal processing for which smoothing out a signal with a top-hat kernel as wide as $T$ strongly suppress signals narrower than $T$. As a consequence, the smoothed signal will contain only signals wider than $T$, so that subtracting it from the actual signal is the same as saving only contributions from frequencies higher than $1/T$. As it will be discussed in a more formal way in the next sections, \tams demonstrated to effectively work for point-like and quasi point-like gamma-ray sources, because in these cases a suitable time interval around the source can be excluded from the integration. The time average alongside the excluded region is quite a robust estimation of the exposure, i.e. - through a simple scaling with the average trigger rate - of the \CR background.

The attention on \tams recently grew up because of their application to detect medium scale excesses on top of the large scale \CR anisotropy. The Milagro collaboration firstly tried to ``adapt'' the Direct Integration method for studies on small to intermediate scales ($10\deg$-$30\deg$) \citep{MilagroMSA}. The attempt came from the assertion that averaging for a time interval $T$ corresponds to make the analysis insensitive to structures wider than $15\deg\,T/1\textrm{\ hr}$ in Right Ascension.
Afterwards, other experiments applied \tams for anisotropy studies, either to estimate the over all exposure or to focus the analysis on a certain angular scale \citep{SK2007, FermiElectrons, Icecube2011, ARGOMSA}. In some cases, the property of filtering out larger structures became the main reason why \tams were used, although no detailed discussion was ever made on the potential biases of these techniques in filtering the large-scale structures.

This paper collects a series of simple calculations and observations on the filtering properties of \tams. As they are applied in a variety of experiments having different operating modes, sky-coverage and trigger rate stability, a general treatment of the matter is impossible. When a specific experimental layout had to be accounted for, the authors made use of a virtual EAS array similar to the ARGO-YBJ experiment \citep{argomoon}, whereas to discuss a likely case of underlying large scale structure the model of the large scale anisotropy of \CRs as given in \citep{Tibet2007} has been used. 

Statistical effects were not considered, i.e. no Poissonian fluctuations around the average event content of each pixel were accounted for. In fact, this contribution is to outline some major potential systematic effects, intrinsic to the application of \tams, regardless if the number of events is sufficient to make them visible or not.

The paper is organized as follows. In the section \ref{sec:exposure} an introduction on \tams as exposure estimation methods is given. The section \ref{sec:RA_normalization} is a brief interlude which demonstrates a consequence of data-normalization along the right ascension to be considered for all further discussions, though mostly affecting the $\ell=1,2,3$ components of the signal. In the section \ref{sec:TAMfiltering} the effect of \tams on the signal to be detected are introduced, mostly for what concerns the reduction of the intensity and the appearance of border effects. The section \ref{sec:TAM_MSA} finally provides quantitative information on the residual contribution from filtered components and the signal distortion due to the method. Some conclusive remarks are given in the last section.
%
%
%
\section{Exposure calculation with TAMs}
\label{sec:exposure}
%
From the experimental viewpoint, the observation of excess (or deficit) effects at a level of 10$^{-4}$ is a difficult task, because of the intrinsic uncertainty that \CR apparatus have to cope with in estimating the exposure. For EAS arrays the atmosphere is part of the detector itself and data must be handled with care to avoid that a atmospheric change mimics a signal somewhere in the sky. For detectors on board satellite, no atmosphere effects are there but trigger rate variations persist related to changing conditions along the orbit.

In general, assuming there is an isotropic charged \CR flux overwhelming all the other signals, the exposure is estimated by assuming it proportional to the integrated \CR flux. In this way, the exposure estimation problem is posed as a \CR-counting problem.

Hereafter, the number of events collected (or computed) in the solid angle $\de \Omega$ centered around $\Omega=(\theta,\phi)$ in the local frame, in the time interval $[t,t+\de t)$ will be written as:
\begin{equation}
\nonumber
\frac{\de N(\Omega,t)}{\de t}\left(=\frac{\de^2 N(\Omega,t)}{\de\Omega\,\de t}\right)
\end{equation}
to lighten the notation.
\subsection{Point-like and \emph{quasi-}point-like sources}
For point-like or \emph{quasi-}point-like sources, \tams are usually applied to estimate the exposure (i.e. the expected background \CR rate) from a certain direction of the sky. They are an evolution of the elder ``on-off'' method and rely on the assumption that the \CR flux from a given direction $\Omega$ in the local reference frame is practically constant during short time-periods. In other words, the average count from $\Omega=(\theta,\phi)$ during the the interval $T$ is quite a good approximation of the \CR number $N_{cr}$:
\begin{equation}
  \frac{\de N_{b}(\Omega,t)}{\de t}\simeq\frac{\de \widetilde{N}_{b}(\Omega,t)}{\de t}=\left\langle \frac{\de N_{ev}(\Omega,t)}{\de t}\right\rangle_{w,T}
\label{eq:estimatedB}
\end{equation}
where $N_{b}$ is the actual (unknown) background \CR number, $\widetilde{N}_{b}$ is the estimated one, and $N_{ev}$ indicates the number of \emph{measured} events. The average is computed in the time interval $T$ and using the kernel function $w$, so that:
\begin{equation}
  \left\langle \frac{\de N_{ev}(\Omega,t)}{\de t}\right\rangle_{w,T}=\frac{\int_{t-T/2}^{t+T/2}\de \tau\ \frac{\de N_{ev}(\Omega,\tau)}{\de \tau}w(\tau)}{\int_{t-T/2}^{t+T/2}\de \tau\ w(\tau)}
\label{eq:timeav}
\end{equation}
If the source contribution is not excluded, the function $w(\tau)$ in \eqref{eq:timeav} is the trigger rate and accounts for over-all variations in the acquisition regime:
\begin{equation}
\de \tau\ w(\tau)=\de\tau\left[\int_{\rm{FOV}} \de\Omega\ \frac{\de N_{ev}(\Omega,t)}{\de t}\right]_{t=\tau}
\label{eq:wdef}
\end{equation}
The integration is carried out numerically, with the Direct Integration or the Time Swapping method (see the Introduction).

The next sections of this paper will focus on the role of $\de N_{ev}/\de\tau$ in the estimate \eqref{eq:timeav}, as this quantity is the sum of different contributions and the problem of a proper separation of the signal in the angular domain via \tams has to be approached by considering the time properties of $\de N_{ev}/\de\tau$. However, before that, two other aspects of the equations \eqref{eq:estimatedB}-\eqref{eq:timeav} should be made explicit.
\begin{itemize}
\item{Time interval.} The quality of the approximation is related to the difference between $N_{b}(\Omega,t)$ and $\widetilde{N}_{b}(\Omega,t)$ (\ref{eq:estimatedB}), i.e. to how representative the time-average is of each instant \CR-flux. If the time window $T$ is chosen too wide, the geometrical distribution of the \CR arrival directions may significantly change, due to atmospheric effects. Some changes in the detector operating regime may have the same effect of making the $\de N_{ev}/\de \tau$ distribution not uniform.
\item{Source exclusion.} The source contribution should be excluded from the time-average. Mathematically, the weight \eqref{eq:wdef} has to be replaced by $w_{se}(\tau)$:
  \begin{equation}
    w(\tau)\longrightarrow w_{se}(\tau)=
    \left\{
      \begin{array}{ll}
        0&\mbox{ if $\Omega\in D_{src}(\tau)$}\\
        w(\tau)&\mbox{ otherwise}
      \end{array}
    \right. 
    \label{eq:timeav2}
  \end{equation}
  where $D_{src}(\tau)$ indicates a confidence solid angle around the source at the time $\tau$.
\end{itemize}

As far as the time window is concerned, the acquisition of \eas arrays is not stable for periods longer than $2-3$ hrs, as climatic changes affect either the arrival direction distribution of \CR and the detector response to the incoming radiation. There are far minor problems for underground experiments or neutrino observatories, where even longer times are used in the literature (up to 24 hrs \citep{Icecube2011}). Nonetheless, time intervals as short as $4$ hrs or less are used also in some of these cases to extract small-scale signals. Satellite-borne detectors usually are so stable to allow to the experimenter to shuffle events within the whole data-set available (up to few years), so that the analysis dows not suffer the pitfalls described below \citep{FermiElectrons}.

About the source exclusion, the solid angle to be excluded around the source is related to the detector angular resolution. A safe choice might be 2 or 3 times the average angular resolution \emph{plus the source intrinsic extension}. If a $2\deg\mbox{-wide}$ source is observed with an angular resolution of $1\deg$, a safe exclusion region of $6\deg-8\deg$ around it can be set. If the region is populated of other known sources, the definition of the exclusion region has to be obviously adapted.

For all experiments surveying the sky, the \fov does not coincide with the portion of the celestial sphere to be investigated. They exploit the rotation of the laboratory frame with respect to the sidereal frame to get the project coverage. In this sense, all time-spans may be translated into angular intervals measured in the sidereal frame. If the laboratory rotates  around the Earth axis (ground-based experiments), time intervals are \ra intervals. Depending on the rotation of the laboratory frame in the sidereal frame, 1 hr may correspond to $\sim15\deg$ in \ra for a ground-based detector or $\sim240\deg$ in the orbit panle of a low Earth circular orbit satellite. For the IceCube detector, at the South Pole, the sky portion observed is always the same and time-flow simply brings a rotation with respect to the celestial coordinates. For ground-based detectors $2-3$ hrs correspond to $30\deg-45\deg$ and enough statistics is left to allow the source exclusion ($\sim 50-80\%$ of the events inside the time window $T$ can be used). In the literature, typical values are found to be $T=2$~hrs for the time interval and $\Delta=6\deg$ for the exclusion region width \citep{Fleysher, argomrk421}. 
\subsection{Wider structures}
If wider structures are considered, the two conditions of the previous section cannot be fulfilled \emph{at the same time}. In fact, the off-source integration interval becomes narrower than the source extension, thus making the on/off source event ratio too high and introducing large fluctuations in the exposure estimation.

This is true for a number of structures having physical extension wider than few degrees. For instance, when experiments like Milagro or ARGO-YBJ measure the diffuse emission from the Galactic Plane, the source exclusion region is usually a $\pm 5\deg$ galactic latitude belt around the plane. Studies of systematics are performed by extending the region up to $\pm10\deg$, obtaining a non-negligible contribution to the uncertainty ($\sim10\%$ \citep{MilagroDiffuse}). The $\pm5\deg$ choice gives less fluctuations but probably still includes some signal events in the background estimation. On the contrary, the $\pm10\deg$ is a safer choice for what concerns the source exclusion, at expense of the statistics\footnote{It should be noticed that $\pm10\deg$ in Galactic Latitude corresponds to a varying \ra interval, as the Galactic plane is not oriented along the celestial equator.}. Quoting this effect as a source of systematics is still acceptable because the experiments do not have the sensitivity to extend the measurement up to $10\deg-15\deg$ from the Galactic Plane. Perhaps next-generation experiments will have it and it will not be possible to exclude the whole region of interest when applying \tams.

A similar point holds for the MSA regions, often wider than $20\deg$, for which the source exclusion is not applicable.

In these cases, the signal intensity is reduced by a factor $\rho$ depending on the signal and the background morphology, as well as on the time-window chosen to apply the \tam. For uniform background, uniform source with extension $T_S$ in local hour angle and time-window $T$ it holds $\rho=1-T_S/T$.
\section{TAMs and LSA}
\label{sec:RA_normalization}
Before coping with the filtering properties of \tams, we discuss here the effect of \tams on the measurement of the \lsa of cosmic rays. Actually, no modern experiment but IceCube used \tams to estimate the exposure \citep{Icecube2011} for all-scale analysis, because it would mean to average along $24$ hrs and to face all the issues of detector stability addressed in the previous section. Nonetheless, the result reported here is valid also for all the other measurements of the \CR anisotropy, e.g. ``equi-zenith'' \citep{Tibet2005} or ``forward-backward'' \citep{MilagroLSA} or else. In fact, a common device to bypass the ignorance of the absolute detection efficiency as a function of the arrival zenith angle (i.e. of the declination), is to set the average flux of cosmic rays detected in a certain zenith (declination) belt to a certain value, the same for all different belts. In other words, deviations from the isotropy are not measured with respect to the average over the whole sky observed, as to do that the efficiency of the detector as a function of the zenith must be properly accounted for. Conversely, the reference average is computed along each zenith belt.

We show here that this solution introduces a degeneracy in the measurement of the anisotropy, i.e. $m=0$ components of the signal are suppressed.

If the signal is looked at as a distribution $f(\theta,\phi)$ on the sphere, the act of normalizing the average content of each declination belt to zero can be written as the operator $S$:
\begin{equation}
  \nonumber
  f(\theta,\phi) \xrightarrow{S}\,f^\prime(\theta,\phi)=f(\theta,\phi)-\frac{1}{2\pi}\int_0^{2\pi}\de\phi\,f(\theta,\phi) 
\end{equation}
where the $f^\prime$ distribution is the measured one, which differs from the ``true'' $f$ for the average $\langle f\rangle_\theta=1/2\int_0^{2\pi}\de\phi\,f(\theta,\phi)$. We can consider the spherical harmonics expansion of the $f$ distribution:
\begin{equation}
  \nonumber
  f(\theta,\phi) =\sum_{\ell=0}^{\infty}\sum_{m=-\ell}^{\ell}a^{\ell}_m\,Y_{\ell}^m(\theta,\phi)
\end{equation}
and considering in a closer detail the effect of the average on the signal. In fact:
\begin{equation}
 \nonumber 
 \int_0^{2\pi}\de\phi\,Y_{\ell}^m(\theta,\phi)=\begin{cases}0&\mbox{ if }m\neq0\\Y_{\ell}^m(\theta,\phi)&\mbox{ if }m=0\end{cases}
\end{equation}
Using the last result, $f^\prime$ can be rewritten as:
\begin{equation}
  f^\prime(\theta,\phi)=f(\theta,\phi)-\sum_{\ell=0}^{\infty}a^{\ell}_0\,Y_{\ell}^0(\theta,\phi)
\end{equation}
where the degeneracy is made explicit. In fact, all terms with $m=0$ are suppressed by the experimental technique applied, what is more important as the multipole order $\ell$ gets lower.

If the sky is only partially observed, further effects arise due to the non-uniform exposure. In fact, if the number of events strongly depends on the declination or other preferred directions, significant deviations from isotropy might be observed only in certain regions of the field of view.
 
\begin{figure}
\centering
\subfigure[]{\includegraphics[angle=90,width=0.45\textwidth]{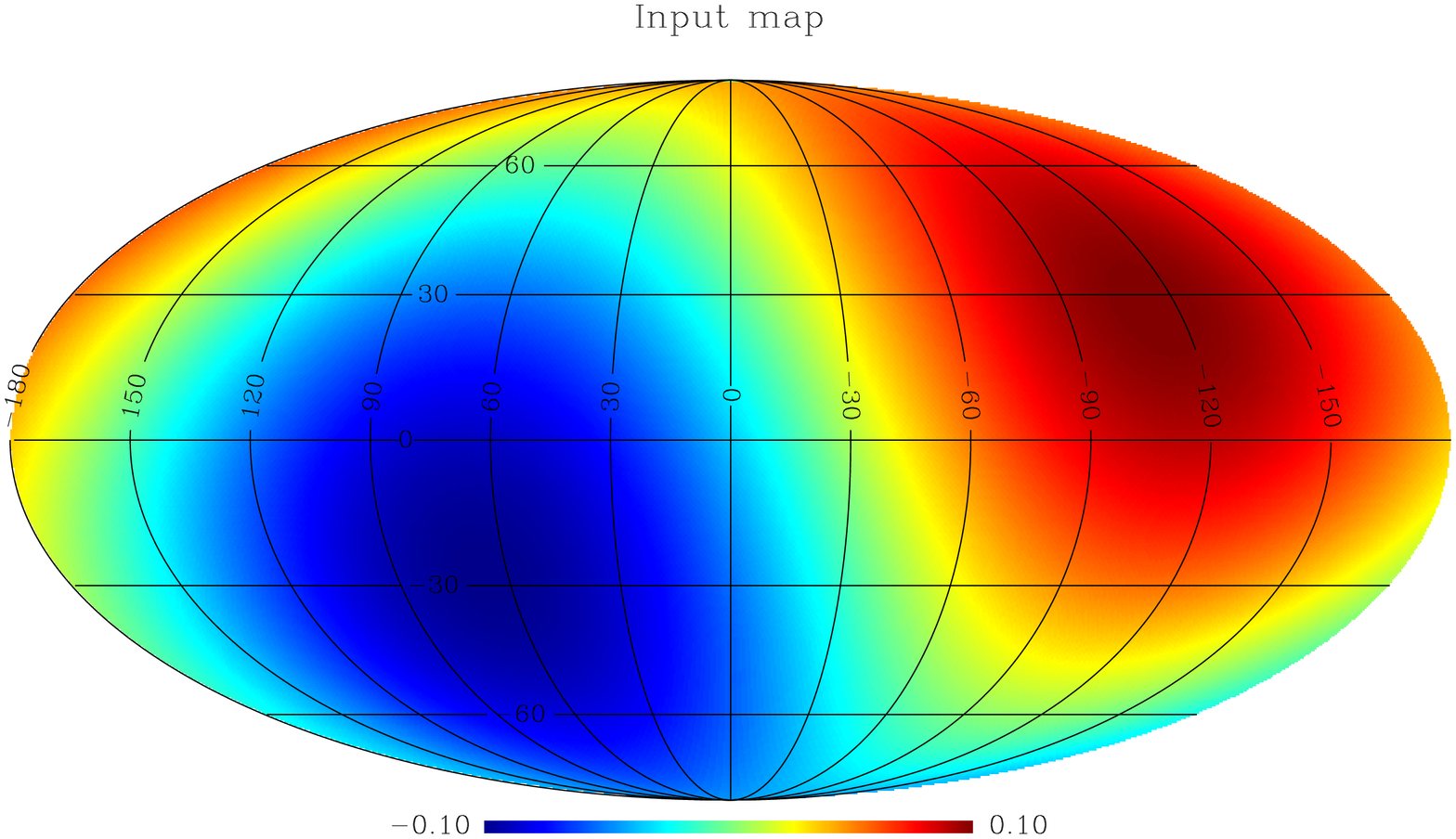}\label{fig:D_input}}
\subfigure[]{\includegraphics[angle=90,width=0.45\textwidth]{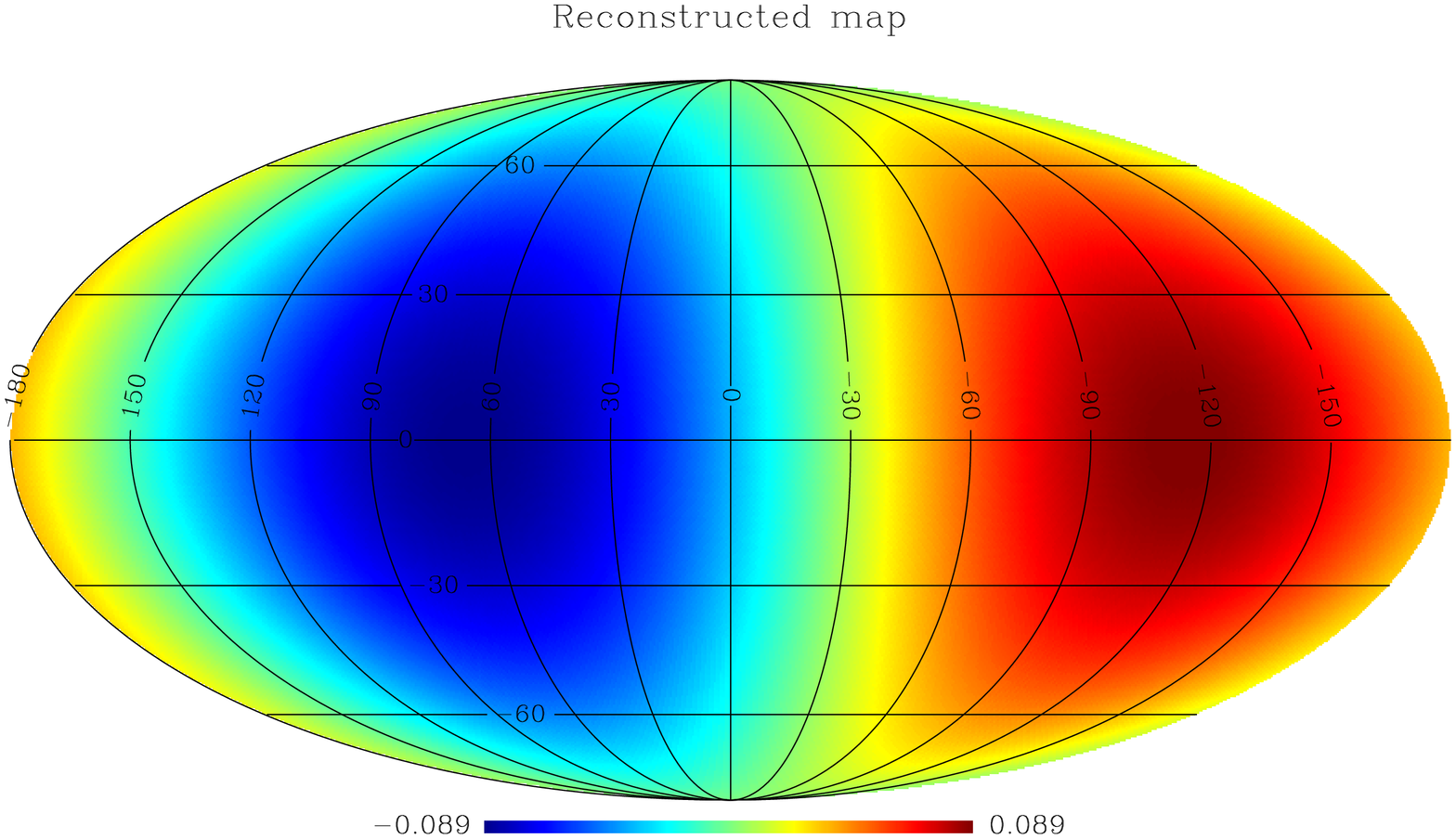}\label{fig:D_reco}}
\subfigure[]{\includegraphics[angle=90,width=0.45\textwidth]{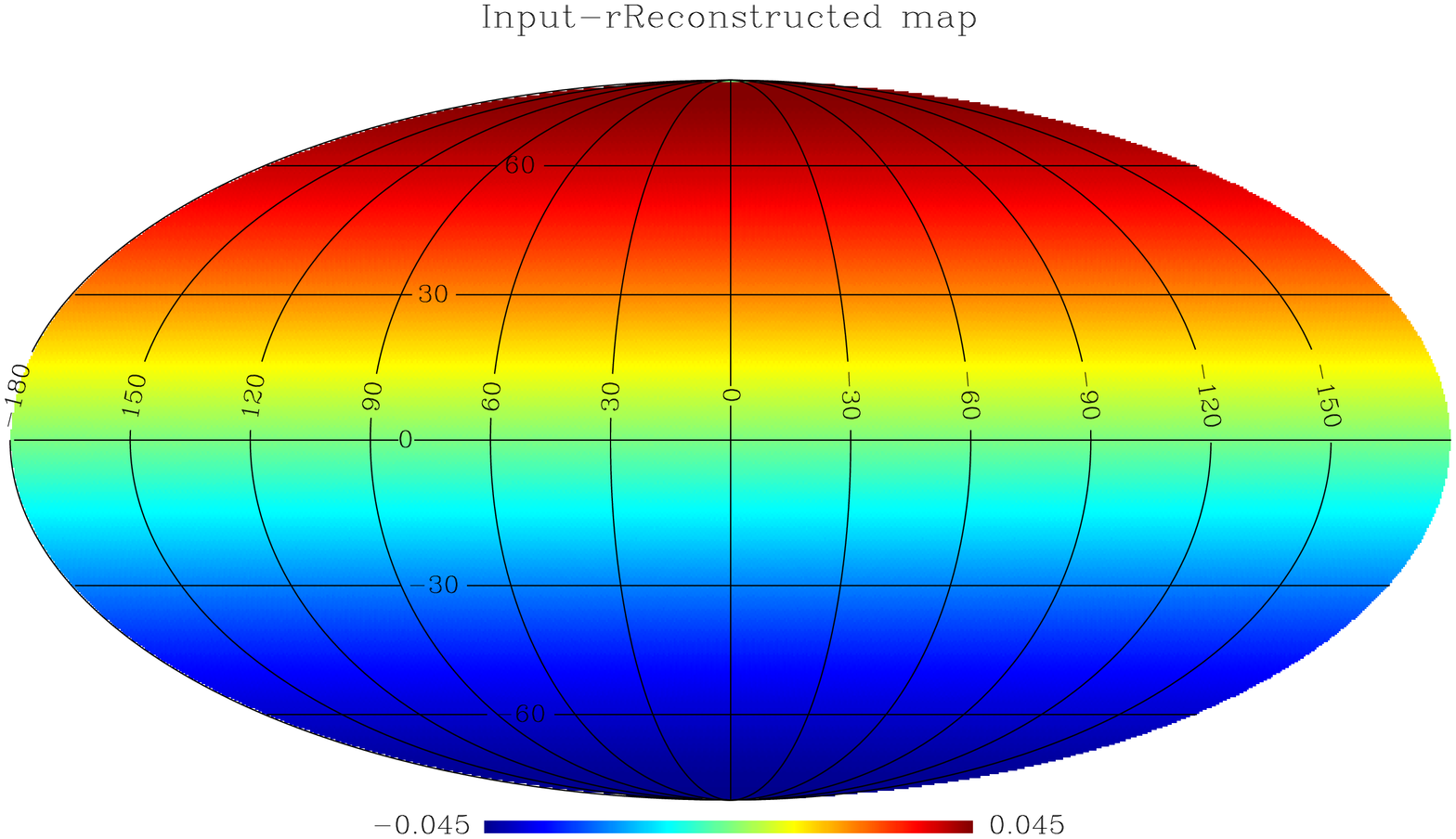}\label{fig:D_residual}}
\caption[Effect of the \ra normalization on a dipole signal.]{Effect of the \ra normalization on a dipole signal. Figure \subref{fig:D_input} represents the input dipole signal, as intense as $0.1$. The dipole vector points at $(\theta=63\deg,\phi=243\deg)$. Figure \subref{fig:D_reco} represents the dipole reconstructed with methods normalizing the average content in each \dec belt to zero. The dipole vector points at $(\theta=90\deg,\phi=243\deg)$ and the intensity is $0.089$. Figure  \subref{fig:D_residual} represents the difference between the input map and the reconstructed one, which turns out to be a dipole as intense as $0.045$, pointing at $\theta=0\deg$. Notice that $0.089^2+0.045^2=0.1^2$.}
\label{fig:D}
\end{figure}

\begin{figure}
\centering
\subfigure[]{\includegraphics[angle=90,width=0.45\textwidth]{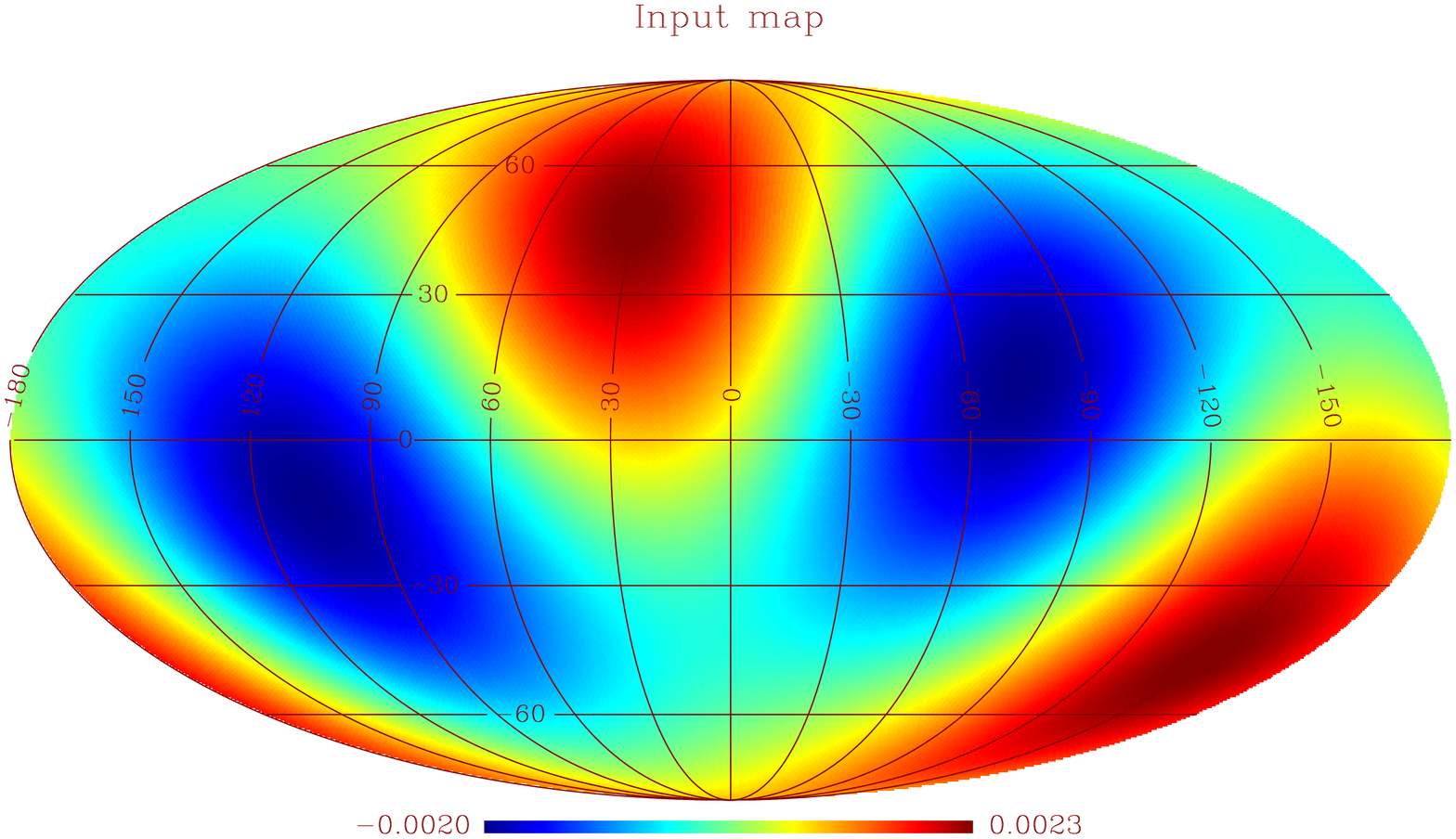}\label{fig:Q_input}}
\subfigure[]{\includegraphics[angle=90,width=0.45\textwidth]{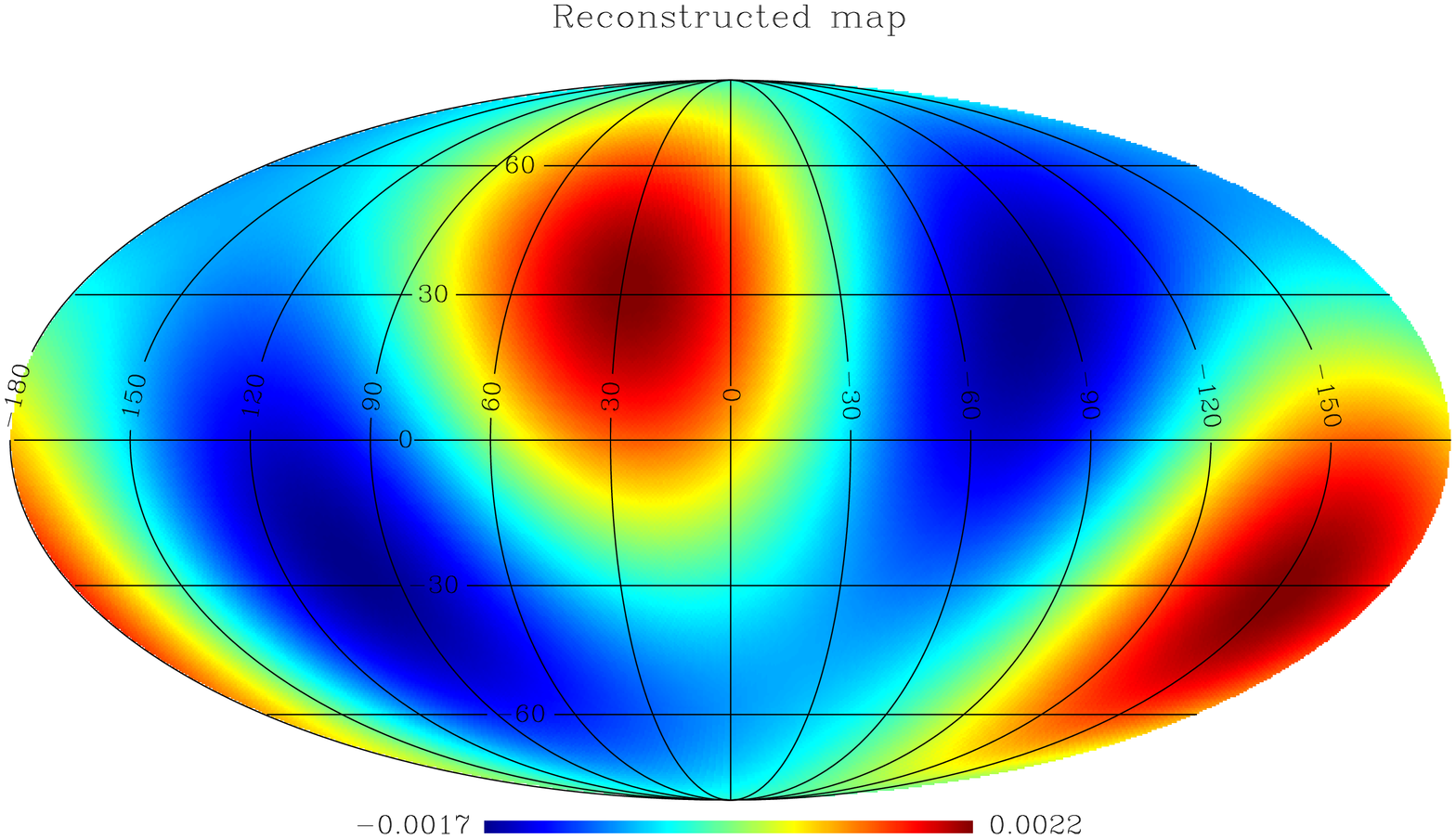}\label{fig:Q_reco}}
\subfigure[]{\includegraphics[angle=90,width=0.45\textwidth]{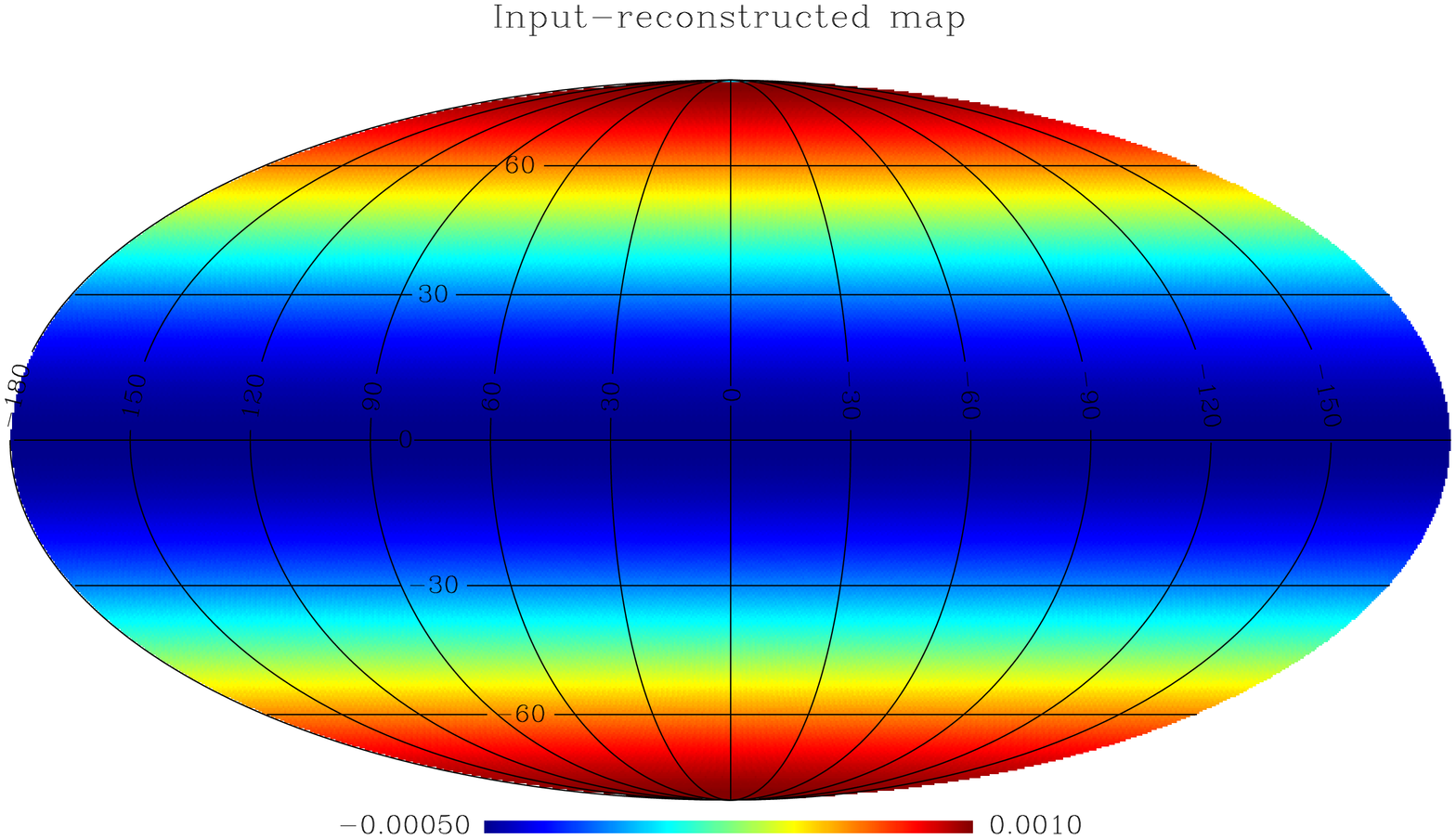}\label{fig:Q_residual}}
\caption[Effect of the \ra normalization on a quadrupole signal.]{Effect of the \ra normalization on a quadrupole signal. Figure \subref{fig:Q_input} represents the input quadrupole signal. Figure \subref{fig:Q_reco} represents the quadrupole reconstructed with methods normalizing the average content in each \dec belt to zero. Figure  \subref{fig:Q_residual} represents the difference between the input map and the reconstructed one, which turns out to be proportional to the $Y_0^2(\theta,\phi)$ function.}
\label{fig:Q}
\end{figure}

A representation of the effect just described is given for $\ell=1,2$ in figures \ref{fig:D}-\ref{fig:Q}.
\section{Filtering properties of TAMs}
\label{sec:TAMfiltering}
As time is a synonym for \ra, \tams average signals along the \ra direction, i.e. they enjoy the property of filtering out large scale contributions to the signal.

\begin{figure*}[!htpb]
\centering
\subfigure[]{\includegraphics[width=0.8\textwidth]{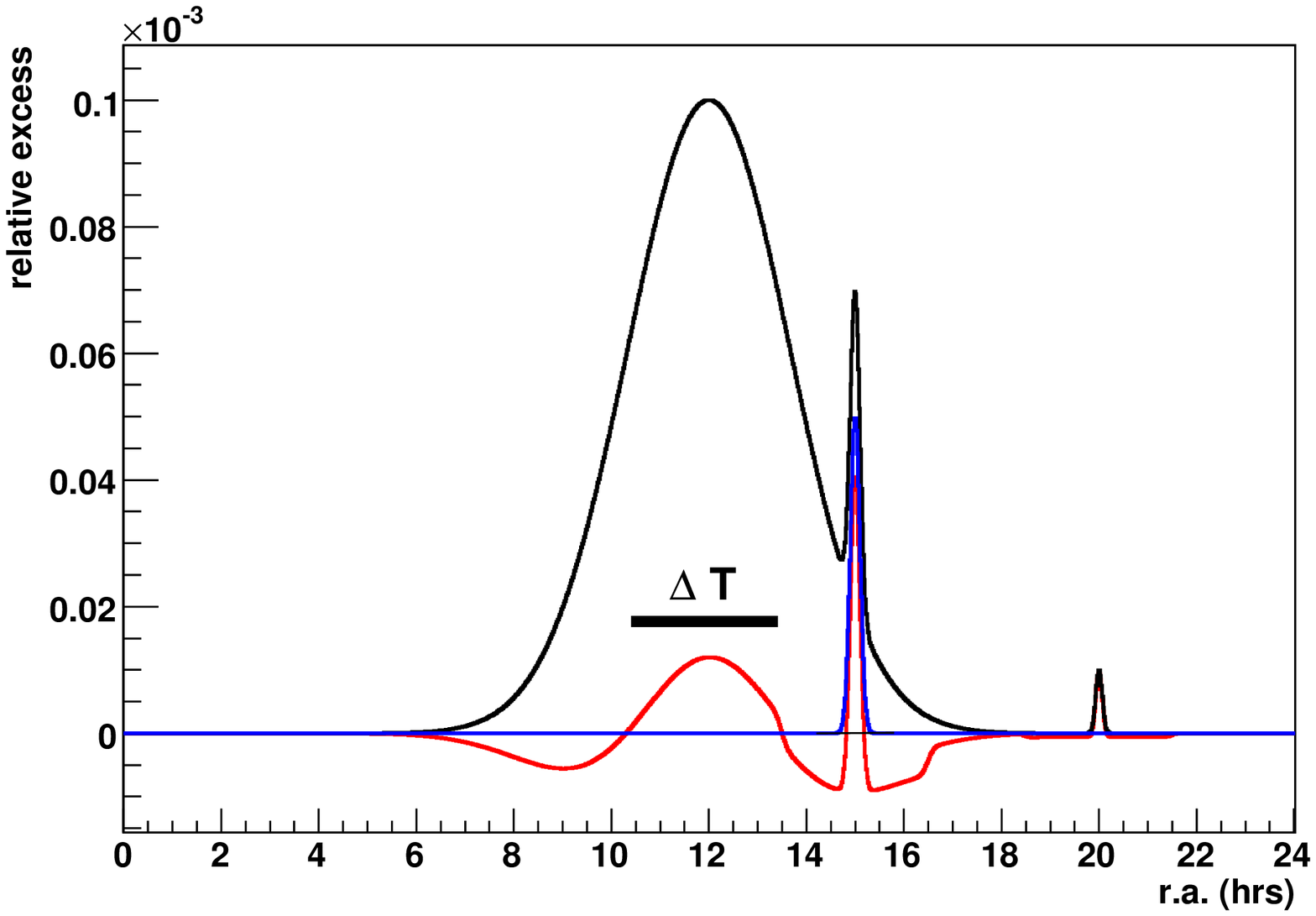}\label{fig:TS_smearingEffect}
}
\subfigure[]{\includegraphics[width=0.4\textwidth]{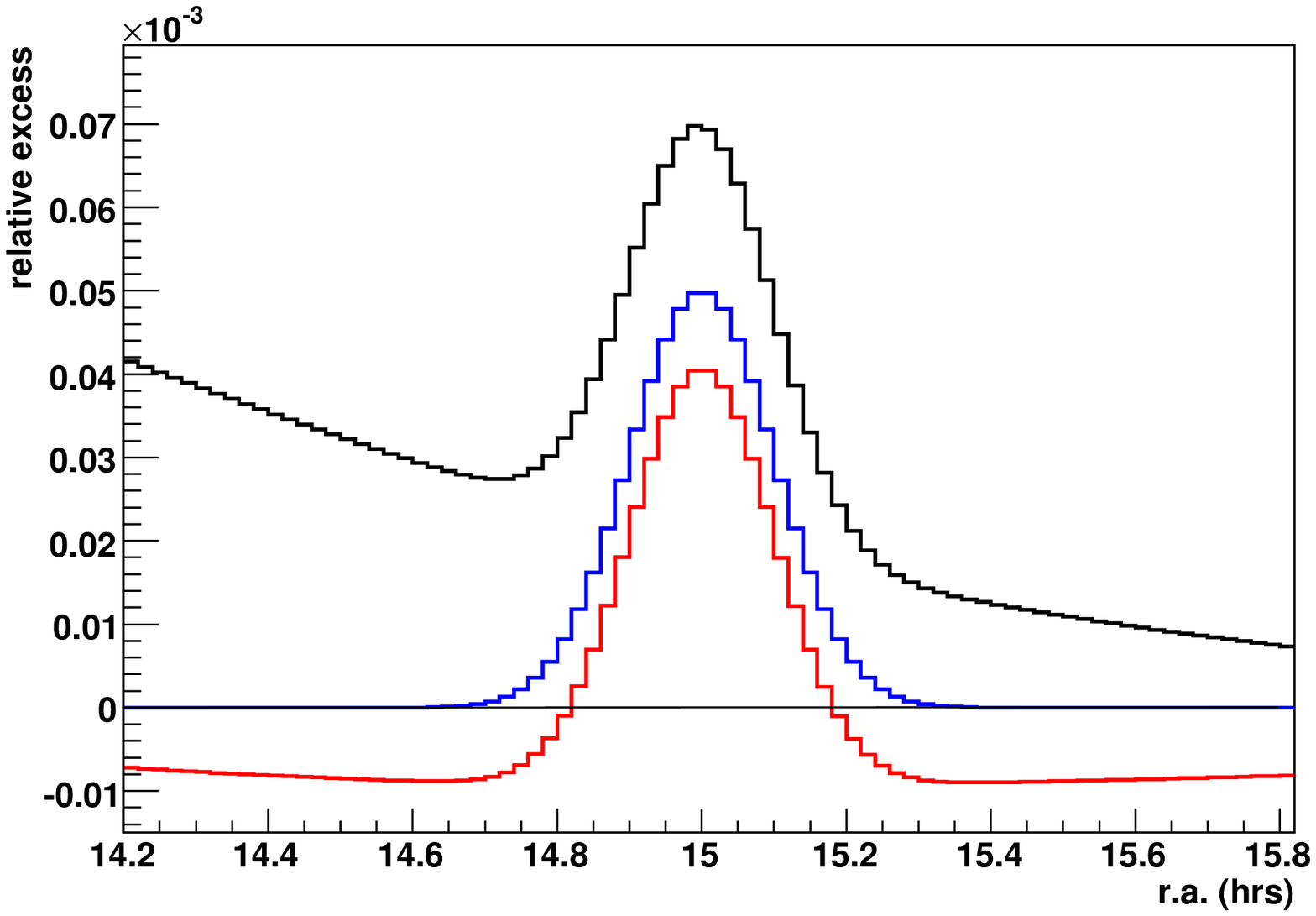}\label{fig:TS_smearingEffect_zoom1}
}
\subfigure[]{\includegraphics[width=0.4\textwidth]{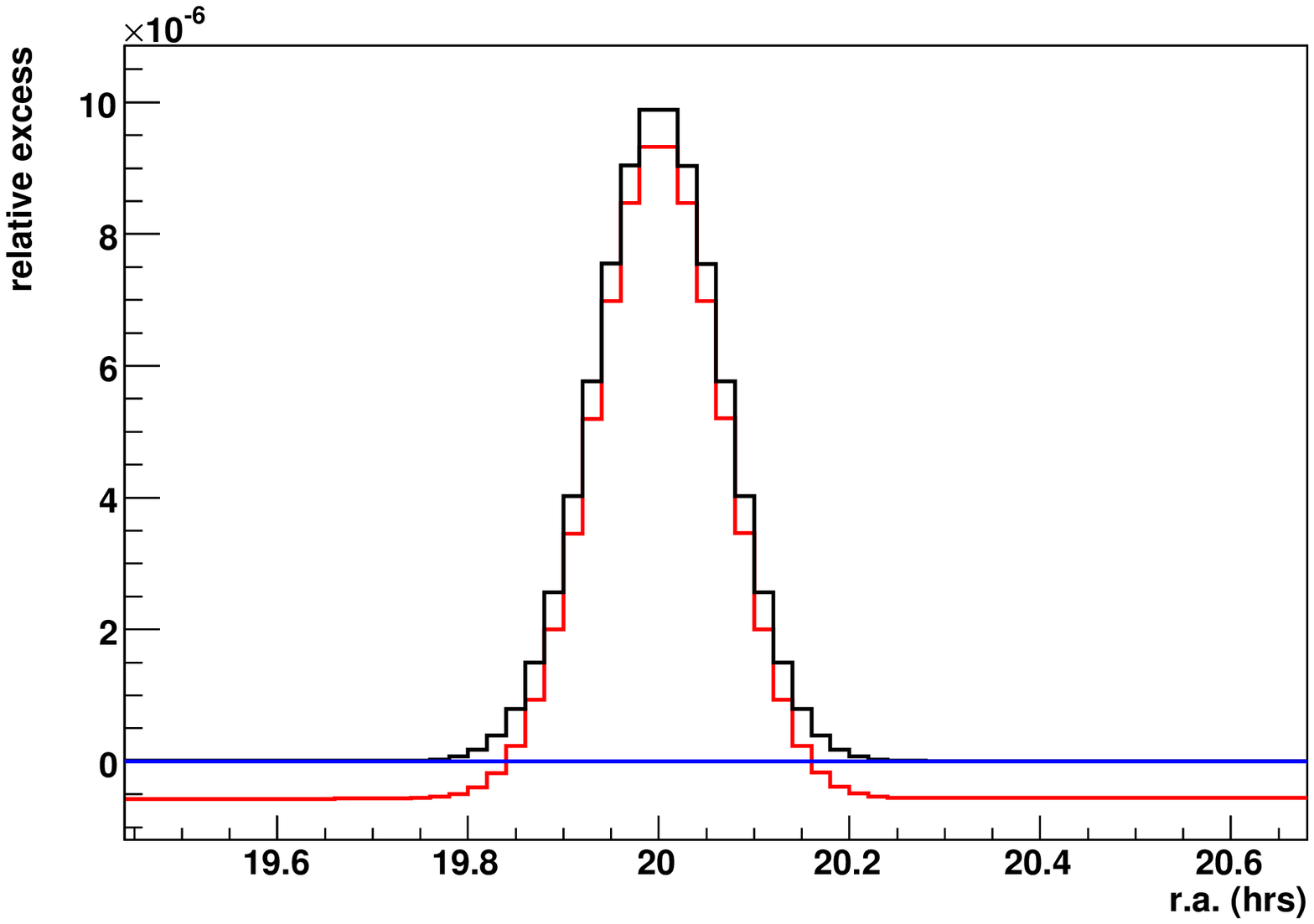}\label{fig:TS_smearingEffect_zoom2}
}
\caption[Toy Monte Carlo on the time-average background methods.]{Simulation of the effect of the time average on the estimate of the intensity of signals of different angular extensions. The central highest signal peaks around 12 hrs, another signal is visible in the spike around 15 hrs and is represented by the blue curve in the zoom \subref{fig:TS_smearingEffect_zoom1}. One more signal, around 20 hrs, is represented also in the zoom \subref{fig:TS_smearingEffect_zoom2}. {\em Black curve:} sum of all the signals. {\em Red curve:} the signal of the black curve as it would appear after the application of a 3 hrs-wide \tam. {\em Blue curve:} input signal centered at $\ra=15\mbox{ hrs}$.}
\label{fig:th1}
\end{figure*}

If we have a data series $x_k\ (k=1,2,\dots,\mathcal{N})$ and we compute for each point the average:
\begin{equation}
\xi_n=\frac{1}{N}\ \sum_{k=n-N/2}^{n+N/2}x_k\ \ \ (N\leq\mathcal{N}) 
\end{equation}
$k=k\pm N \mbox{ if $k<1$ or $k>N$)}$ then the difference $x-\xi$ will maintain intact all structures narrower than $N$, whereas all features much wider than $N$ will be suppressed. In Fig. \ref{fig:th1} we show the results of a toy numerical estimation of the time-average effect on the signal intensity estimation for different angular scales.
The red curve clearly shows that the average along $\Delta T$ preserves signals on narrower angular scales and strongly reduces wider contributions. 
The Fig. \ref{fig:th1} triggers some other considerations. Firstly the excess (or the deficit) is observed as less intense than it really is. This bias can be avoided by excluding the source region, what is impossible for structures wider than half the time-window extension. A second important issue (related to the first one) is that fake deficit zones are rendered around true excesses and, vice-versa, fake excesses are seen around true deficits. The importance of this problem lays in the absolute lack of knowledge \emph{a-priori} of which signal is true and which is not. The problem is present mostly for structures coming from reducing wider features than for really-narrow signals. If the actual signal is well separated in the harmonic space from the wider structure to be suppressed, it may be possible to observe it also without any filter, i.e. just by considering the \ra distribution like in \citep{MilagroMSA}. On the other hand, if some hypothesis can be made (from the literature or from independent data about the detector exposure), the effect of such underlying larger scale signals can be easily estimated and quoted as systematic uncertainty of the final measurement.
\section{TAMs and MSA}
\label{sec:TAM_MSA}
The study of the \msa in the arrival distribution of \CR can be approached with different methods.

Likely the most orthodox way is to evaluate the exposure with techniques sensitive to any angular scale and then apply the spherical harmonics analysis to filter out the signal. The use of the $a_{\ell m}$ coefficients prevents any contamination from harmonic regions other than those selected with the $(\ell,m)$ numbers and allows to define the degree of anisotropy in a (mathematically) robust way.

Another approach, still starting from an all-scale-sensitive estimation of the exposure, is to estimate the dipole and quadrupole components of the measured \CR distribution, to subtract them from the experimental distribution and to focus on the residuals at scales less than $90\deg$. For such narrower signals, the analysis is carried out in the real domain.

These two methods enjoy the uncontested feature of properly filtering the signal from scales larger or equal to $90\deg$, although some problems are there due to the partial coverage of the sphere by the experimental data. In fact, either the $a_{\ell m}$ expansion and the dipole-quadrupole determination are achieved with fit procedures over the whole sphere: that part of the sky which is empty of data has to be suitably masked and the lack of information unavoidably reflects upon the error associated to the intensity and the structure position on the sphere.

Moreover, both the $a_{\ell m}$ and the dipole-quadrupole methods rely on an estimation of the exposure all over the angular scales, what implies that they can filter out the \lsa \emph{only} if it is properly detected. If the all-scale analysis revealed some systematics for the \lsa, it would be difficult to make sense from an $a_{\ell m}$ expansion of such a signal.

On the other hand, we already hinted that the application of time-average techniques to get the \msa signal would introduce systematics on the flux estimation, as well as the bias of filtering \emph{only} along the \ra direction.

Nevertheless, two arguments are in favor of these techniques:
\begin{enumerate}
\item they do not inherit systematics from effects present below the angular scale they are set to filter out. In this sense, no Compton-Getting interference is expected, neither influence or artifacts induced by large scale atmospheric effects, which instead were demonstrated to be relevant for the \lsa analysis; In fact, systematics introduced in misinterpreting the detector performance usually affects the whole sky, hardly being responsible for localized features; 
\item the amplitude of the systematics described above, i.e. the residuals from the \lsa structures, can be evaluated with Monte Carlo simulations if independent measurements or robust models are available.
\end{enumerate}
\subsection{Residuals of underlying large-scale structures}
Before considering how \tams reconstruct extended signals like the MSA, i.e. which effects the analysis technique has on the intensity and the shape of the signal under observation, it is worth to investigate the residual effect of the underlying large-scale structures, which are strongly suppressed by the time-average.
\vspace{3mm}\\
\textbf{Signal gradient along the RA direction}\\
\label{sec:ragrad}
The result of the \tam depends on the signal which it is applied to, so that no prediction is possible if the signal \emph{in toto} is not considered. Nonetheless, it is convenient to focus on two characteristics of the signal separately, i.e. the extension and the gradient along the RA direction. It is easy to assess that the gradient of the estimated background is proportional to the difference of the \emph{signal} gradient at the boundaries of the time average window:
$$\frac{\de}{\de t}\left\langle \frac{\de N_{ev}(\Omega,t)}{\de t}\right\rangle_{w,T}\simeq$$
\begin{equation}
  \frac{k}{T}\left(\left.\frac{\de N_{ev}(\Omega,\tau)}{\de \tau}\right|_{\tau=t+T/2}-\left.\frac{\de N_{ev}(\Omega,\tau)}{\de \tau}\right|_{\tau=t-T/2}\right)
  \label{eq:RA_grad}
\end{equation}
The constant $k$ depends on the kernel function used\footnote{It must be symmetric in time, i.e. $w(\tau)=w(-\tau)$.} In the numerical implementation, if a ``top-hat'' kernel is used, it turns out to be $k=1$. The equality does not hold because of the denominator in the equation \ref{eq:timeav}, introducing second order corrections in $\de N_{ev}(\Omega,t)/\de t$.
The equation \ref{eq:RA_grad} can be figured out by thinking of a discrete implementation of the \tam, where a moving time window passes from the $i\mbox{-th}$ time-bin to the $(i+1)\mbox{-th}$. The content of the bin centered at $t_{i+1}+T/2$ is included in the background estimation in place of the content of the bin centered at $t_{i}-T/2$.

\begin{figure}[!htbp]
\centering
\includegraphics[width=0.95\linewidth]{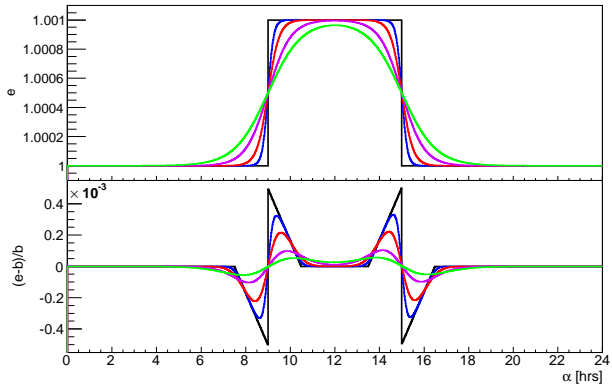}
\caption{Toy-calculation to highlight the relevance of the gradient along the RA direction of the signal under study. \emph{Top panel:} a 6-hrs wide signal as intense as $10^{-3}$ with respect to the underlying flat background was simulated to rise up with different slope; \emph{bottom panel:} the signal as reconstructed after the \tam-calculated background subtraction (time-window: 3 hrs).}
\label{fig:RA_gradient}
\end{figure}

A simple representation of this effect is given in figure \ref{fig:RA_gradient}. The top panel represent a large scale excess as intense as $I=10^{-3}$ with respect to the isotropic \CR flux, drawn according to the equation:
$$\frac{\de N_{ev}(\Omega,t)}{\de t}=$$
\begin{equation}
\nonumber
\frac{I}{2}\left[\tanh\left(\frac{t-\alpha_0+\Delta\alpha/2}{w}\right)-\tanh\left(\frac{t-\alpha_0-\Delta\alpha/2}{w}\right)\right]
\end{equation}
The signal center is fixed at $\alpha_0=12\mbox{ hrs}$, the width at $\Delta\alpha=6\mbox{ hrs}$ and the signal gradient constant $1/w$ is changed from $\infty$ to $1/1.5\mbox{ hrs}^{-1}$ (black to green curves). The bottom panel reports how a \tam (top-hat kernel, time-window 3 hrs) would filter each signal of the panel above. Residuals of the large scale signal remain, whose intensity strongly depends on the signal gradient. For a non-physical signal like the black one ($w=0$), border effects as intense as half the input signal are visible. These effects are reduced below $10\%$ the input intensity if the signal gradient along \ra is less than $1\mbox{ hr}^{-1}$. It can be noticed that residuals are both of positive and negative nature.
The maximum intensity of the residuals and their extension (intended as the width of the intervals wherein they are always above or below $5\%$ the input signal intensity) are reported in the table \ref{tab:RA_gradient}. It can be noticed that the width is never above $3\mbox{ hrs}$, i.e. the time-window used in the analysis.
\begin{table}[!htbp]
  \centering
  \begin{tabular}{c||c|c}
    {\bf Signal gradient}&{\bf Amplitude of}&{\bf Width of}\\
    {\bf ${\rm (hrs^{-1})}$}&{\bf residuals${\rm (\times 10^{-4})}$}&{\bf residuals} (hrs)\\
    \hline
    $\infty$&$5.0$&1.5\\
    $1/0.25$&$3.3$&1.7\\
    $1/0.5$&$2.2$&2.2\\
    $1$&$1.0$&3.0\\
    $1/1.5$&$0.5$&3.0
  \end{tabular}
  \caption{Intensity and extension of residuals reported in the bottom panel of figure \ref{fig:RA_gradient}.}
  \label{tab:RA_gradient}
\end{table}
\vspace{3mm}\\
\textbf{The effect of the LSA on the time-average exposure estimation}\\
In the last section it was shown how important is the gradient of the large scale signal intended to be filtered out with \tams. Results were given for a toy-calculation with the purpose of enhancing potential biases of the analysis.

The aim of this section is to evaluate the effect of the residual contribution of a large scale signal whose nature is closer to reality. Once again, the result was achieved with numerical calculations, and the algorithm applied is described in the following.

\paragraph{\lsa parameterization.} To avoid to introduce any circular bias, we used the parameterization of the \lsa given by the Tibet-AS$\gamma$ collaboration in \citep{amenomori10}. The authors model their observation with two structures, a Global and a Medium Anisotropy. The former signal is what is commonly referred to as \lsa, whereas the second one lays on smaller scales and is part of the signal that the methods discussed in this paper are tuned for. It is worth noticing here that the best fit of the model \citep{amenomori10}, is given after the normalization of data along each declination band (see the section \ref{sec:RA_normalization}).
\paragraph{Exposure simulation.} The detector exposure was simulated by using the local \CR distribution obtained for a flat ground-based detector, with a standard atmosphere absorption model ($dN/d\theta=I_0\exp(-k/\cos\theta)$, with $k=4.8$). After time discretization, the local \CR distribution was computed for each time bin in the sidereal day and transposed in equatorial coordinates. The exposure map showed the characteristic feature of a maximum at a \dec few degrees above the experiment latitude ($30\deg\mbox{ N}$), fading away at the field of view \dec limits . As the calculus was performed with events arriving within $\theta=50\deg$, the reference \dec band for this analysis is $-20\deg\leftrightarrow 80\deg$.
\paragraph{Event simulation.} As already said, in order to exclude uncertainties due to statistics, the \emph{event map} was not filled by following a Poissonian distribution, but using the average value expected from the numerical integration of the local distribution function. This is the reason why no fluctuations are visible in the \ra profile. The \emph{actual background map} was obtained with the same solution: to avoid any fluctuations, any pixel was filled with the product of the exposure times the total number of detected events.
\paragraph{Estimated background.} The estimated background map was obtained from the event map, i.e. from the sky picture \emph{containing} the \lsa signal. The content of each pixel was replaced with the average content of all pixel less distant than $T/2$. The \tam was repeated for all values of $T$ from $1$ to $24$ hrs. Results for 1, 2, 3, 4, 6, 12 and 24 hrs will be reported only.

The Healpix ``ring'' pixelization scheme was used \citep{healpix}.

\vspace{3 mm}
Anticipating the section \ref{sec:declination_effect}, we notice that the angular distance between pixels is the same all over the sphere, but the \ra distance increases when the pixel under consideration are close to the poles. Consequently, as the interval for the average is set in the \ra space, the computation at low \dec values is carried out on more pixels than at higher \dec. The methods becomes ineffective when the number of pixels at a certain \dec is such that the average along a certain $\Delta T$ is the pixel itself, making the background equal to the signal. The effect is small for \dec $\delta<45\deg$ and $T>3$~hrs~$=60\deg$, but is important above $\delta=70\deg$, mostly for $T\leq 2$~hrs. 

\begin{figure*}[!htbp]
\centering
\subfigure[]{\includegraphics[width=0.45\textwidth]{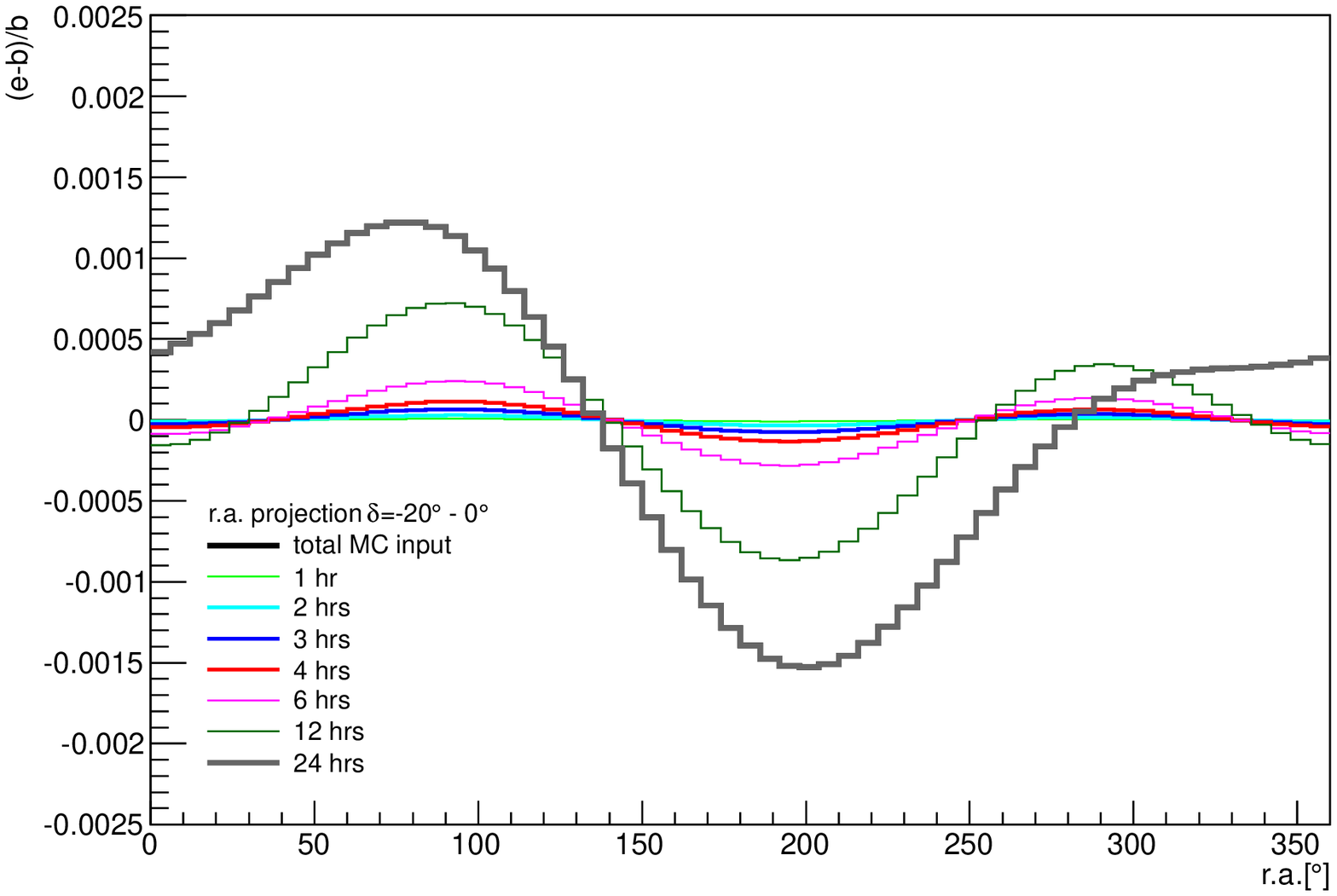}\label{fig:-20+0}}
\subfigure[]{\includegraphics[width=0.45\textwidth]{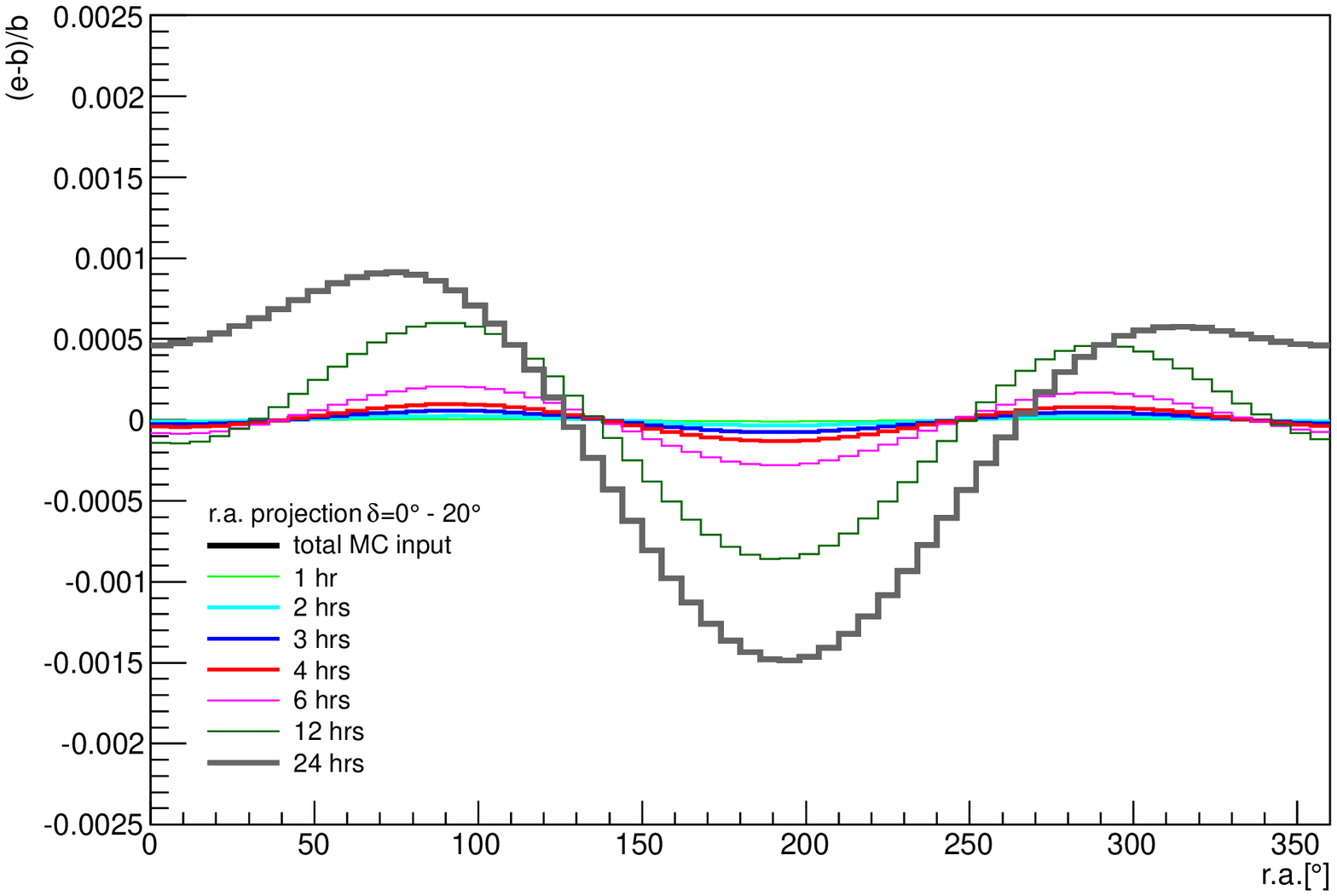}\label{fig:0+20}}
\subfigure[]{\includegraphics[width=0.45\textwidth]{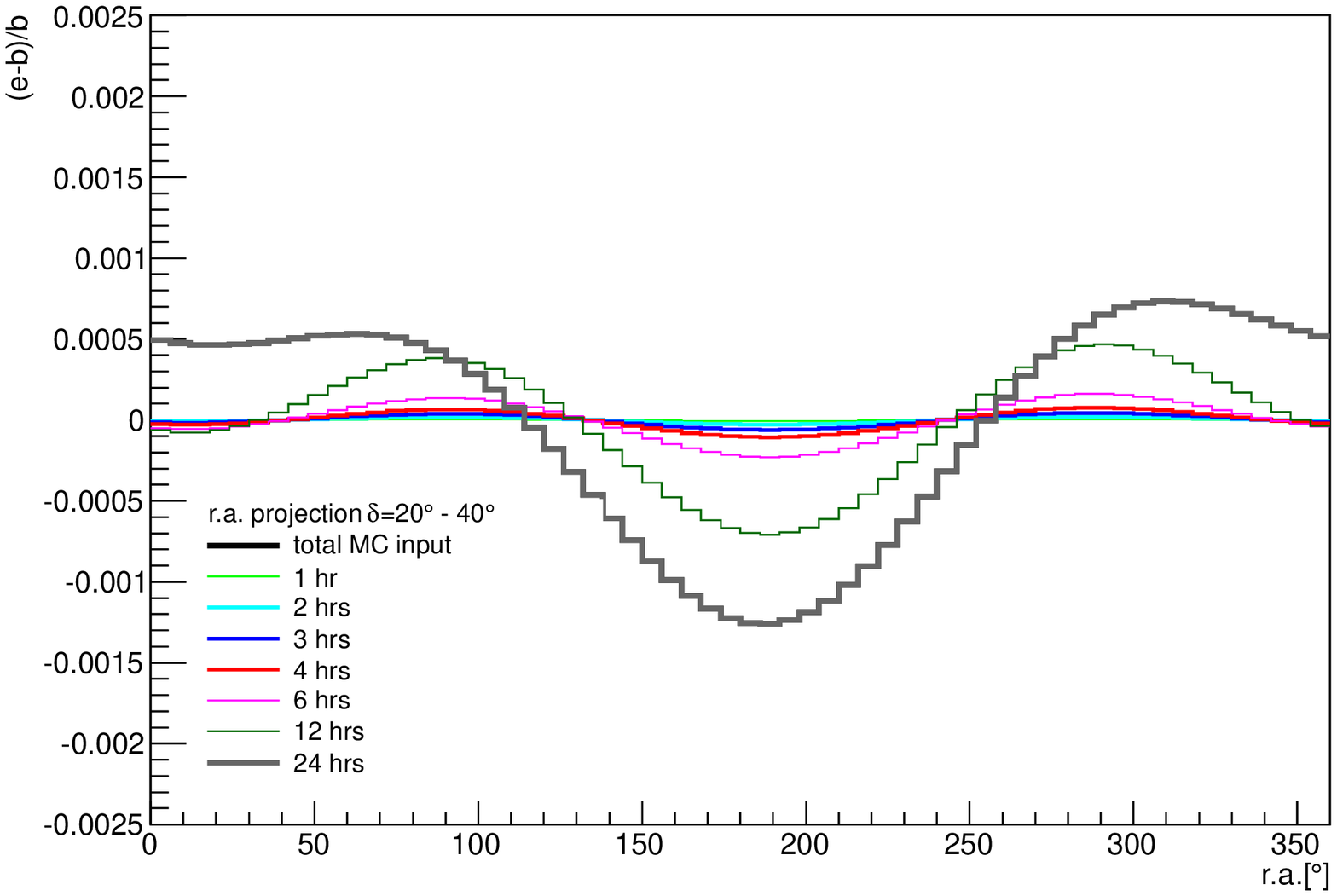}\label{fig:20+40}}
\subfigure[]{\includegraphics[width=0.45\textwidth]{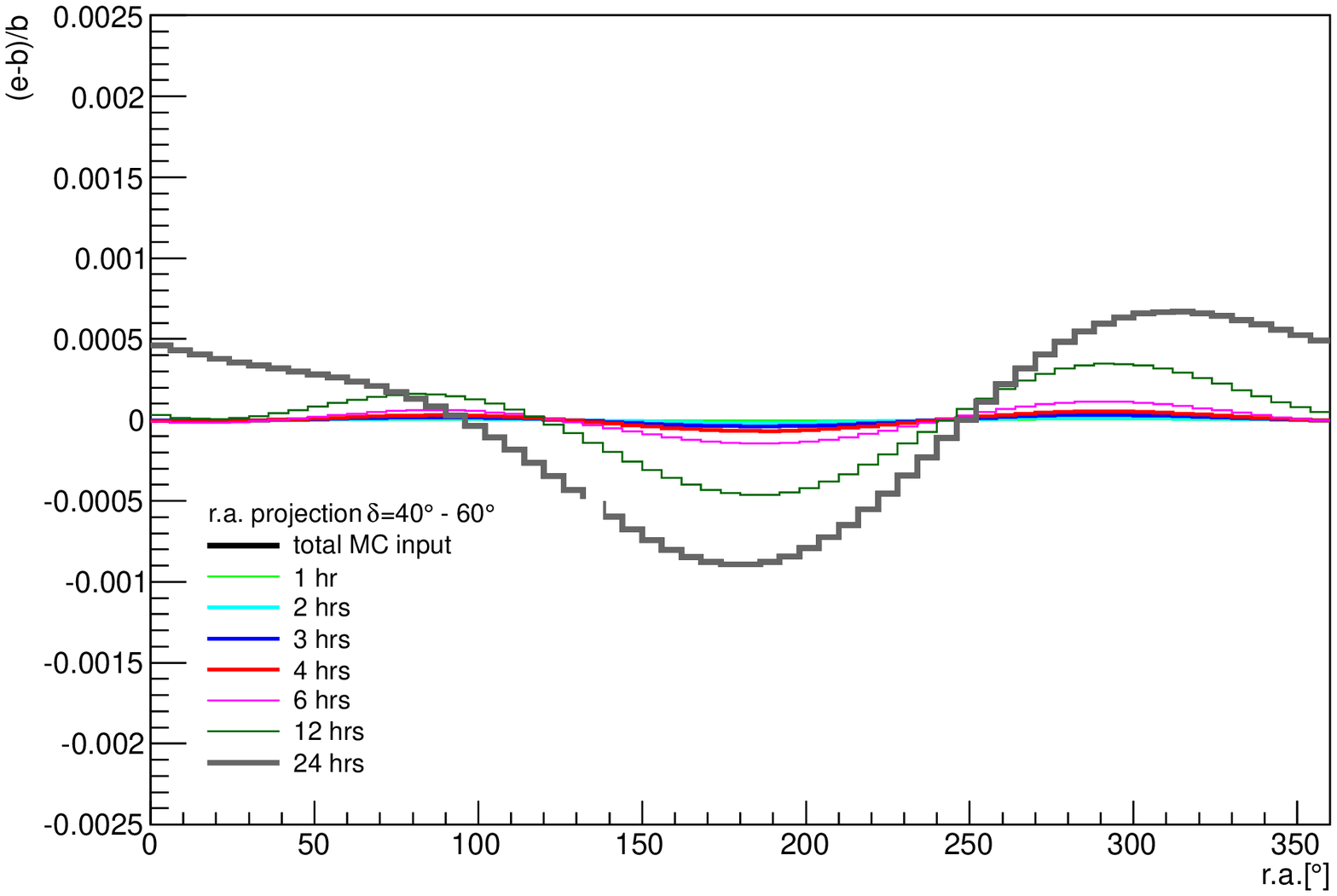}\label{fig:40+60}}
\subfigure[]{\includegraphics[width=0.45\textwidth]{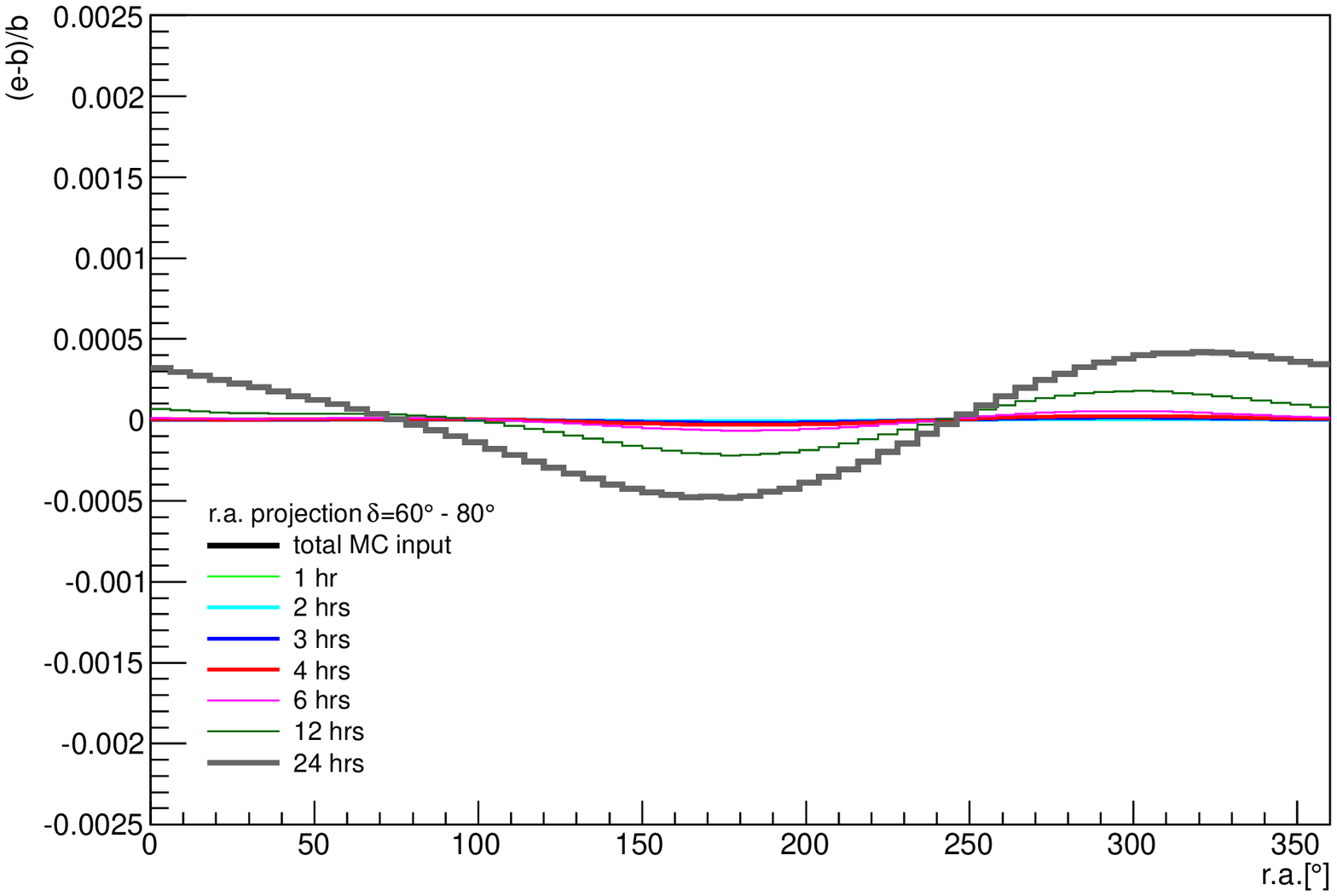}\label{fig:60+80}}
\caption{\lsa simulation for different \dec bands: \subref{fig:-20+0} $-20\deg\leftrightarrow 0\deg$; \subref{fig:0+20} $0\deg\leftrightarrow 20\deg$; \subref{fig:20+40} $20\deg\leftrightarrow 40\deg$; \subref{fig:40+60} $40\deg\leftrightarrow 60\deg$; \subref{fig:60+80} $60\deg\leftrightarrow 80\deg$. Different curves represent different time-average windows (see the legend for details).}
\label{fig:lsa}
\end{figure*}

Figure \ref{fig:lsa} reports the result of the calculus for different \dec bands. The plots were obtained by projecting data of the event map ($\de N_{ev}/\de t$ abbreviated with $e$ in the figure) and the estimated background map ($\de \tilde{N}_{b}/\de t\rightarrow b$) in the declination interval indicated, then by calculating for each bin the ratio $(e-b)/b$. In every plot, the black curve represents the input \lsa signal after the \ra normalization. As no detector-induced effects nor statistical fluctuations are considered, the gray curve perfectly fits with it, representing the signal as it would be observed with an all-scale sensitive \tam ($T=24\mbox{ hrs}$). The other curves represent what remains of the \lsa structure when the $T\mbox{-wide}$ filter is applied. It can be appreciated how the \lsa is practically suppressed for $T\leq 4$~hrs.
The table \ref{tab:lsaeff} reports the absolute maximum deviation of the \lsa-induced signal from the zero-reference-value as a function of the \dec band and the time-average window, in units of $10^{-4}$. It is worth noticing that the maximum effect occurs at the maxima of the \lsa, so that for medium scale structures observed in other regions than those maxima, such an effect is far less than reported in the table \ref{tab:lsaeff}. 
\begin{table}[!htbp]
  \centering
  \begin{tabular}{c||c|c|c|c|c}
    {\bf \dec}&\multicolumn{5}{|c}{{\bf Systematic}}\\
    {\bf band}&\multicolumn{5}{|c}{{\bf error }$\mathbf{(\times 10^{-4})}$}\\
    \hline
    &1 hr&2 hrs&3 hrs&4 hrs&6 hrs\\
    \hline
    $-20\deg\leftrightarrow 0\deg$&0.07&0.29&0.7&1.1&2.4\\
    $0\deg\leftrightarrow 20\deg$&0.06&0.26&0.6&1.0&2.1\\
    $20\deg\leftrightarrow 40\deg$&0.05&0.20&0.4&0.8&1.6\\
    $40\deg\leftrightarrow 60\deg$&0.04&0.14&0.3&0.5&1.1\\
    $60\deg\leftrightarrow 80\deg$&0.016&0.06&0.14&0.25&0.5
  \end{tabular}
  \caption{Systematic error induced by the time-average filtering method used in the analysis. The \lsa  was parameterized according to the ``Global anisotropy'' given in \citep{amenomori10}.}
  \label{tab:lsaeff}
\end{table}
 
Figure \ref{fig:lsa} makes a point in showing how weak the effect of the \lsa is when structures narrower than $45\deg$ are looked for. If the numbers of the table \ref{tab:lsaeff} are considered, together with the typical intensity quoted for the \msa emission ($4\cdot10^{-4}-10^{-3}$ relative to \CR isotropic flux, with $T\leq3\mbox{ hrs}$ \citep{MilagroMSA,ARGOMSA,Icecube2011}), systematic residuals from the \lsa after the \tam are proved to be less than $20\%$. 
\subsection{Reconstruction of MSA structures}
As the signal to be resolved is not excluded in the background computation, distortions will appear in its reconstruction. From previous section it should be clear that the effect of the \tam depends on the actual shape of the event distribution, whose composition is not known \emph{a-priori}, i.e. nobody knows what is signal and what is background. After considering how the underlying wider structures affect the analysis, we describe here the relation between the actual and the reconstructed signal. The results are obtained by considering a top-hat one-dimensional signal to which \tam with different time windows are applied. The background is assumed to be constant, what is equivalent to assuming that the larger scale structures are well separated in the harmonic space from the \msa. In this sense, it should be noticed that the calculation is the same of that reported in the study of the \ra gradient, with the important difference that the ratio of the signal width to the time window was greater than 1 there, but it is smaller here.

\begin{figure}[!htbp]
\centering
\includegraphics[width=0.95\linewidth]{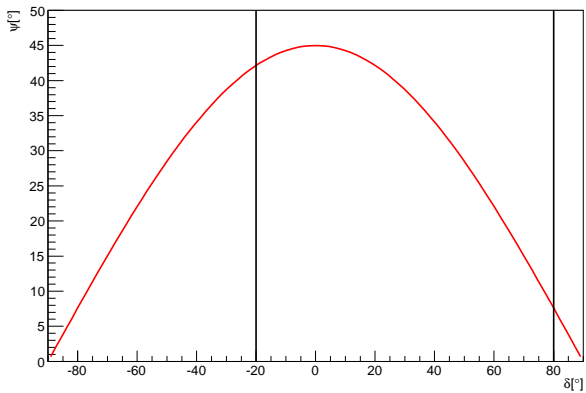}
\caption{Angular distance between two points on the sphere having RA coordinates shifted $45\deg$ from each other, as a function of the declination. The vertical lines enclose the declination region for which many computations of this paper were made.}
\label{fig:psi_VS_dec}
\end{figure}

The issue of the actual angular size of the signal and that of the reduction of the intensity are addressed.
\vspace{3mm}\\
\textbf{Signal extension and declination}\\
\label{sec:declination_effect}
If it is true that time is the same of RA, so that time-average corresponds to RA-average, it does not mean that the filtering properties of \tams are the same all over the sphere. In fact, the angle $\psi$ between two events having coordinates $(\alpha_1,\delta)$ and $(\alpha_2,\delta)$ depends on $\delta$:
\begin{equation}
\cos\psi=\sin^2\delta+\cos^2\delta\,\cos(\alpha_1-\alpha_2)
\label{eq:decl_dependence}
\end{equation}
The equation \ref{eq:decl_dependence} clearly shows the dependence of $\psi$ on $\delta$ (as expected, $\psi=0\deg$ if $\delta=90\deg$). As a consequence, the effect of any filter working in the RA space will be different according to the declination band: a top-hat filter $45\deg$ wide in RA, corresponds to a top-hat filter $31.4\deg$ wide. A representation of the equation \ref{eq:decl_dependence} for $\alpha_1-\alpha_2=45\deg$ is given in figure \ref{fig:psi_VS_dec}.

This is the reason why \tams tend to return structures more and more narrow as declination bands farther from $\delta=0\deg$ are considered.
\vspace{3mm}\\
\textbf{Signal reduction}\\
We consider a signal with intensity $\de N_s/\de t$ and width $w$ (in \ra), above a flat background with intensity $\de N_b/\de t$. It is analyzed with a \tam with time window $T$. The (measured) event content is indicated with $\de N_e/\de t$. The reconstructed signal $a^{\prime}$
\begin{equation}
\nonumber
a^\prime=\frac{\de N_e/\de t\, -\, \de N_{b^\prime}/\de t}{\de N_{b^\prime}/\de t}  
\end{equation}
can be compared with the actual one
\begin{equation}
\nonumber
a=\frac{\de N_e/\de t\, -\, \de N_{b}/\de t}{\de N_{b}/\de t}=\frac{\de N_s/\de t}{\de N_{b}/\de t}
\end{equation}
by studying the ratio $\alpha=a^{\prime}/a$. The ratio will depend on the ratio $\rho=w/T$ and on $a$ itself. It is easy to demonstrate that for $0<\rho\leq 1$ and $\rho a\ll 1$ (conditions fulfilled for the measured \msa intensity) it holds:
\begin{equation}
  \alpha\simeq(1-\rho)(1+\rho\,a)\simeq1-\rho
  \label{eq:sig_reduction}
\end{equation}
i.e. the reduction of the signal intensity depends linearly on the ratio $\rho$.

The equation \ref{eq:sig_reduction} makes explicit one of the major uncertainty induced by \tams: since the signal width is something to be determined, the choice of the time window to be used has to be externally directed (i.e. after experimental trials or other measurements from literature). It is worth recalling that the relation \ref{eq:sig_reduction} is obtained for a particular (and non-physical) case and it gives an upper limit of the dependence of the signal reduction on the quantities $\rho$ and $\alpha$.
%
\section*{Conclusions}
This paper is intended to contribute to the study of the analysis methods implemented to observe extended signals in the cosmic ray arrival direction distribution. In particular, it focuses on the time-average based methods applied when the source cannot be excluded, and it points out how they undoubtedly enjoy properties of filtering in the \ra direction, at expense of important spurious effects introduced in the signal reconstruction, both for what concerns its intensity and its shape. Experiments like Milagro, IceCube and \argo made use of these methods in the last decade \citep{MilagroMSA,ARGOMSA,Icecube2011}.

On the important point of potential residual effects coming from larger signals, supposed to be filtered out, it has been shown that they depend on the gradient of such signals, rather than on their intensity, what makes the bias from the known underlying \lsa really small.

Known distortions of the reconstructed signal were analyzed, giving numerical information about the intensity reduction, the presence of border effects (deficits around excesses and vice-versa) and a sort of ``reshaping'' due to the filter acting only along the \ra direction.

We conclude that the detection of medium and small structures with \tams may hardly mimic fake signals more intense than $2\,10^{-4}$ with respect to the average isotropic \CR flux. Nonetheless, if more detailed studies are attempted, like energy spectrum (i.e. signal intensity) or morphology, care is needed in considering systematic effects introduced by \tams. In the near future, experiments \citep{Icecube2011,HAWC,Lhaaso} will have the sensitivity to go below the $10^{-4}$ level, what will make the choice of \tams less and less effective.

In this sense, medium and small-scale \CR anisotropy are best searched for if a full-scale analysis is applied, with standard spherical harmonics or wavelet techniques \citep{needlet}, allowing the experimenter to focus on well-defined regions in the harmonic space. The possibility of this approach is related to the capability of controlling the detector exposure down to the level of the signal to be observed, i.e. accounting for detector and environment effects as precisely as $10^{-4}-10^{-3}$. Due to this reason, time-average methods, if carefully implemented, can still be considered a good compromise for current experiments, which are controlled to this level but do not have yet the sensitivity to detect signals as low as $10^{-5}$.
%
%
%

\end{document}